\documentclass[
  ,final            % use final for the camera ready runs
    ,sort&compress
  ]
  {aipproc}

\layoutstyle{6x9}

\begin{document}

\title{Threshold Perspectives on Meson Production}

\author{M. Wolke \thanks{email: \tt{magnus.wolke@tsl.uu.se, m.wolkel@fz-juelich.de}}\ }{
  address={The Svedberg Laboratory, Uppsala University, Box 533, 75121 
    Uppsala, Sweden\setcounter{footnote}{0}
    \renewcommand{\thefootnote}{\fnsymbol{footnote}}\footnotemark},
%  altaddress={Institut f\"ur Kernphysik, Forschungszentrum J\"ulich, 
%           52425 J\"ulich, Germany}
}

\begin{abstract}
Studies of meson production in nucleon--nucleon collisions at threshold are 
characterised by few degrees of freedom in a configuration of well defined 
initial and final states with a transition governed by short range dynamics.
Effects from low--energy scattering in the exit channel are inherent to the 
data and probe the interaction in baryon--meson and meson--meson systems 
otherwise difficult to access.

From dedicated experiments at the present generation of cooler rings precise 
data are becoming available on differential and eventually spin observables 
allowing detailed comparisons between complementary final states.
To discuss physics implications of generic and specific properties, recent 
experimental results on meson production in proton--proton scattering obtained 
at CELSIUS and COSY serve as a guideline.
\end{abstract}

\maketitle

\renewcommand{\thefootnote}{\fnsymbol{footnote}}
\footnotetext{on leave from Institut f\"ur 
    Kernphysik, Forschungszentrum J\"ulich, 52425 J\"ulich, Germany}
%%%%%%%%%%%%%%%%%%%%%%%%%%%%%%%%%%%%%%%%%%%%
%% MAINMATTER
%%%%%%%%%%%%%%%%%%%%%%%%%%%%%%%%%%%%%%%%%%%%

\setcounter{footnote}{0}
\renewcommand{\thefootnote}{\arabic{footnote}}
\section{Introduction}
High precision data from the present generation of cooler rings, IUCF, CELSIUS, 
and COSY, have contributed significantly over the last decade to our present 
knowledge and understanding of threshold meson production (for a recent review 
see \cite{Moskal:2002jm}).

Due to the high momentum transfers required to create a meson or mesonic 
system in production experiments close to threshold the short range 
part of the interaction is probed.
In nucleon--nucleon scattering, for mesons in the mass range up to 
$1\,\mbox{GeV}/\mbox{c}^2$ distances from $0.53\,\mbox{fm}$ ($\pi^0$) down to 
less than $0.2\,\mbox{fm}$ ($\phi$) are involved.
At such short distances it is a priori not clear, whether the relevant degrees 
of freedom are still baryons and mesons, or rather quarks and gluons.
As there is no well defined boundary, one goal of the threshold production 
approach is to explore the limits in momentum transfer for a consistent 
description using hadronic meson exchange models.
Within this framework, questions concerning both the underlying meson exchange 
contributions and especially the role of intermediate baryon resonances have 
to be answered.

Another aspect which enriches the field of study arises from the low relative 
centre--of--mass velocities of the ejectiles:
Effects of low energy scattering are inherent to the observables due to strong 
final state interactions (FSI) within the baryon--baryon, baryon--meson, and 
meson--meson subsystems.
In case of short--lived particles, low energy scattering potentials are 
otherwise difficult or impossible to study directly.

\section{Dynamics of the Two Pion System}
In $\gamma$ and $\pi$ induced double pion production on the nucleon the 
excitation of the $N^*(1440)$ $\mbox{P}_{11}$ resonance followed by its decay 
to the $N \sigma$ channel, i.e.\ $N^*(1440) \rightarrow 
p {(\pi\pi)}_{\,I\,=\,l\,=\,0}$, is found to contribute non--negligibly close 
to threshold \cite{Oset:1985wt,Bernard:1995gx,GomezTejedor:1996pe}.
Nucleon--nucleon scattering should provide complementary information, 
eventually on the $\pi \pi$ decay mode of the $N^*(1440)$, which plays an 
important part in understanding the basic structure of the second excited state 
of the nucleon \cite{Morsch:2000xi,Krehl:1999km,Hernandez:2002xk}.

Exclusive CELSIUS data from the PROMICE/WASA setup on the reactions $pp 
\rightarrow p p \pi^+ \pi^-$, $pp \rightarrow p p \pi^0 \pi^0$ and $pp 
\rightarrow p n \pi^+ \pi^0$ 
\cite{Brodowski:2002xw,Johanson:2002hs,Patzold:2003tr} are well described by 
model calculations \cite{Alvarez-Ruso:1998mx}:
For the $\pi^+ \pi^-$ and $\pi^0 \pi^0$ channels, the reaction preferentially 
proceeds close to threshold via heavy meson exchange and excitation of the 
$N^*(1440)$ Roper resonance, with a subsequent pure s--wave decay to the 
$N \sigma$ channel\footnote{For the $p n \pi^+ \pi^0$ final state, this reaction 
mechanism is trivially forbidden by isospin conservation. An underestimation of 
the total cross section data \cite{Johanson:2002hs} by the model predictions 
\cite{Alvarez-Ruso:1998mx} might be explained by the neglect of effects from 
the $p n$ final state interaction in the calculation 
\cite{Alvarez-Ruso:2002pc}.}.
While nonresonant contributions are expected to be small, resonant processes 
with Roper excitation and decay via an intermediate $\Delta$ ($pp \rightarrow 
p N^* \rightarrow p \Delta \pi \rightarrow p p \pi \pi$) and $\Delta \Delta$ 
excitation ($pp \rightarrow \Delta \Delta \rightarrow p \pi p \pi$) are strongly 
momentum dependent and vanish directly at threshold.
Double $\Delta$ excitation, which is expected to dominate at higher excess 
energies beyond $\mbox{Q} = 250\,\mbox{MeV}$ \cite{Alvarez-Ruso:1998mx} involves 
higher angular momenta and consequently strongly anisotropic proton and pion 
angular distributions.
On the other hand, the Roper decay amplitude via an intermediate $\Delta$ 
depends predominantly on a term symmetric in the pion momenta 
(eq.\eqref{eq_roper_decay}), leading to the 
$p {(\pi^+ \pi^-)}_{\,I\,=\,l\,=\,0}$ channel and an interference with the 
direct $N \sigma$ decay.

Experimentally, for the reaction $pp \rightarrow p p \pi^+ \pi^-$ at excess 
energies of $\mbox{Q} = 64.4\,\mbox{MeV}$ and $\mbox{Q} = 75\,\mbox{MeV}$ 
angular distributions give evidence for only s--waves in the final state, in 
line with a dominating $pp \rightarrow p N^* \rightarrow 
p p {(\pi^+ \pi^-)}_{\,I\,=\,l\,=\,0}$ process, with the initial inelastic $pp$ 
collision governed by heavy meson ($\sigma,\rho$) exchange. 
\begin{figure}[hbt]
  \label{fig_pipi}
  \includegraphics[width=\textwidth]{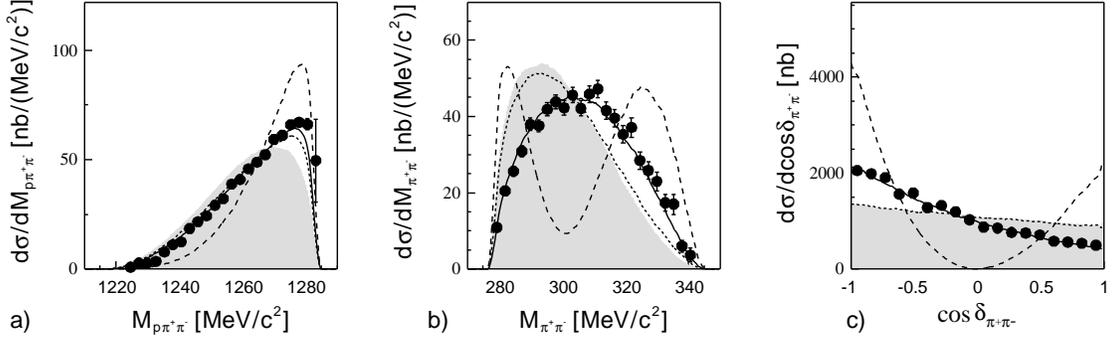}
  \caption{Differential cross sections for the reaction $pp \rightarrow 
     p p \pi^+ \pi^-$ at an excess energy of $\mbox{Q} = 75\,\mbox{MeV}$. 
     Experimental data (solid circles) for invariant mass distributions 
     of the $(p \pi^+ \pi^-)$-- (a) and $(\pi^+ \pi^-)$--subsystems (b), 
     and the $\pi^+ \pi^-$ opening angle (c) are compared to pure phase space 
     (shaded areas) and Monte Carlo simulations for direct decays $N^* 
     \rightarrow N \sigma$ (dotted lines), decays via an intermediate $\Delta$ 
     resonance $N^* \rightarrow \Delta \pi \rightarrow N \sigma$ (dashed lines) 
     and an interference of the two decay routes (solid lines) according to 
     eq.\eqref{eq_roper_decay}. Figures are taken from \cite{Patzold:2003tr}.}
\end{figure}
Roper excitation is disclosed in the $p \pi^+ \pi^-$ invariant mass 
distribution (Fig.\ref{fig_pipi}a), where the data are shifted towards higher 
invariant masses compared to phase space in agreement with resonance excitation 
in the low energy tail of the $N^*(1440)$.
Compared with Monte Carlo simulations including both heavy meson exchange for 
$N^*$ excitation, and $pp$ S--wave final state interaction, but only the direct 
decay $N^* \rightarrow p {(\pi^+ \pi^-)}_{\,I\,=\,l\,=\,0}$ (dotted lines), the 
production process involves additional dynamics, which is apparent from 
discrepancies especially in observables depending on the $\pi$ momentum 
correlation $\vec{\mbox{k}}_{\,1} \cdot \vec{\mbox{k}}_{\,2}$, i.e.\ 
$\pi^+ \pi^-$ invariant mass $\mbox{M}_{\,\pi\pi}$ (Fig.\ref{fig_pipi}b) and 
opening angle $\delta_{\,\pi\pi} = \angle (\vec{\mbox{k}}_{\,1} \cdot 
\vec{\mbox{k}}_{\,2})$ (Fig.\ref{fig_pipi}c).
A good description of the experimental data is achieved including the 
$N^*(1440)$ decay via an intermediate $\Delta$ and its destructive interference 
with the direct decay branch to the $N \sigma$ channel (solid lines) in the 
ansatz for the Roper decay amplitude \cite{Alvarez-Ruso:1998mx}:
\begin{equation}
{\cal{A}} \propto 
1 + c\,\vec{\mbox{k}}_{\,1} \cdot \vec{\mbox{k}}_{\,2} \,
\left(3\mbox{D}_{\Delta^{++}} + \mbox{D}_{\Delta^0}\right),
\label{eq_roper_decay}
\end{equation}
where the first term describes the direct decay, the parameter $c$ adjusts the 
relative strengths of the two decay routes, and $\mbox{D}_{\Delta}$ denote the 
$\Delta$ propagators.
A fit to the data allows to determine the ratio of partial decay widths 
${\cal{R}}(\mbox{M}_{N^*}) = 
\Gamma_{N^* \rightarrow \Delta \pi \rightarrow N \pi \pi} / 
\Gamma_{N^*
 \rightarrow N \sigma}$ at average masses $<\mbox{M}_{N^*}>$ 
corresponding to excess energies $\mbox{Q} = 64.4\,\mbox{MeV}$ and 
$\mbox{Q} = 75\,\mbox{MeV}$ relative to the $\pi^+ \pi^-$ threshold.
The numerical results, ${\cal{R}}(1264) = 0.034 \pm 0.004$ and ${\cal{R}}(1272) 
= 0.054 \pm 0.006$, exhibit the clear dominance of the direct decay to the 
$N \sigma$ channel in the low energy region of the Roper resonance. 
On the other hand they indicate the strong energy dependence of the  
ratio from the momentum dependence in the decay branch via an intermediate 
$\Delta$, which will surpass the direct decay at higher energies 
\cite{Patzold:2003tr}.
A model dependent extrapolation based on the validity of ansatz 
\eqref{eq_roper_decay} leads to ${\cal{R}}(1440) = 3.9 \pm 0.3$ at the nominal 
resonance pole in good agreement with the PDG value of $4 \pm 2$ 
\cite{Hagiwara:2002fs}.

Within the experimental programme to determine the energy dependence of the 
$N^* \rightarrow N \pi \pi$ decay exclusive data (for details see 
\cite{Bashkanov:2003ha}) have been taken simultaneously at the CELSIUS/WASA 
facility on both the $p p \pi^+ \pi^-$ and $p p \pi^0 \pi^0$ final states.
In case of the $\pi^+ \pi^-$ system the preliminary results at an excess energy 
of $\mbox{Q} = 75\,\mbox{MeV}$ are in good agreement with the relative strength 
of the decay routes adjusted to an extrapolated ratio 
${\cal{R}}(1440) = 3$.
However, at slightly higher excess energy ($\mbox{Q} = 127\,\mbox{MeV}$) the 
data might be equally well described by a value ${\cal{R}}(1440) = 1$, which is 
noticeably favoured at both excess energies by the data on $\pi^0 \pi^0$ 
production, indicating distinct underlying dynamics in $\pi^0 \pi^0$ and 
$\pi^+ \pi^-$ production.
One difference becomes obvious from the isospin decomposition of the total 
cross section \cite{Johanson:2002hs}: 
An isospin $\mbox{I} = 1$ amplitude in the $\pi \pi$ system, and accordingly a 
p--wave admixture, is forbidden by symmetry to contribute to the neutral pion 
system in contrast to the charged complement.
A p--wave component was neglected so far in the analysis, since the unpolarized 
angular distributions show no deviation from isotropy.
However, there is evidence for small, but non--negligible analysing powers from 
a first exclusive measurement of $\pi^+ \pi^-$ production with a polarized beam 
at the COSY--TOF facility \cite{Bashkanov:2003ha,Clement:2003ar}, suggesting 
higher partial waves especially in the $\pi \pi$ system.

At higher energies, i.e.\ $\mbox{Q} = 208\,\mbox{MeV}$ and $\mbox{Q} = 
286\,\mbox{MeV}$ with respect to the $\pi^+ \pi^-$ threshold, preliminary data 
for both $\pi^+ \pi^-$ and $\pi^0 \pi^0$ from CELSIUS/WASA rather follow phase 
space than expectations based on a dominating $pp \rightarrow p N^* \rightarrow 
p p \sigma$ reaction mechanism\cite{Bashkanov:2003ha}.
At these energies, the $\Delta \Delta$ excitation process should influence 
observables significantly, and, thus, a phase space behaviour becomes even more 
surprising, unless the $\Delta \Delta$ system is excited in a correlated way.

\section{The Proton--Proton--Eta Final State}
\begin{figure}[hbt]
  \label{fig_eta_etap}
  \includegraphics[height=.40\textheight]{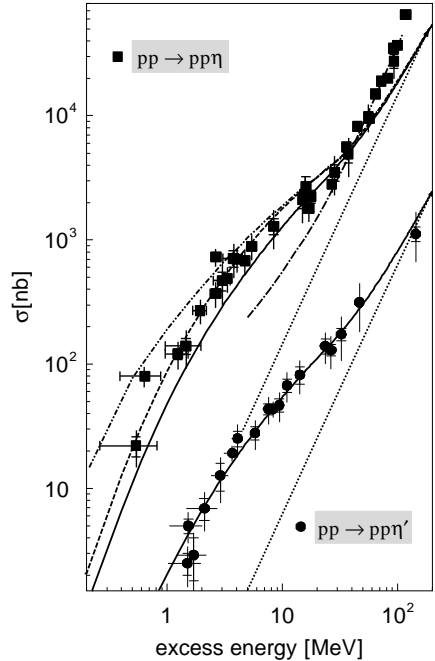}
  \caption{Total cross section data for $\eta$ (squares \cite{Bergdolt:1993xc,
     Chiavassa:1994ru,Calen:1996mn,Calen:1997sf,Hibou:1998de,Smyrski:1999jc}) 
     and $\eta^\prime$ (circles \cite{Hibou:1998de,Moskal:1998pc,Moskal:2000gj,
     Balestra:2000ic,Khoukaz:2001mn1}) production in proton--proton scattering 
     versus excess energy Q \cite{Moskal:2003gt}. 
     In comparison, the energy dependences from a pure phase space behaviour 
     (dotted lines, normalized arbitrarily), from phase space modified by the 
     $\mbox{}^1\mbox{S}_0$ proton--proton FSI including Coulomb interaction 
     (solid lines), and from additionally including the proton--$\eta$ 
     interaction phenomenologically (dashed line), are shown. Meson exchange 
     calculations for $\eta$ production including a P--wave component in the 
     proton--proton system \cite{Nakayama:2003jn} are depicted by the 
     dashed--dotted line, while the dashed--double--dotted line corresponds to 
     the arbitrarily normalized energy dependence from a full three--body 
     treatment of the $pp \eta$ final state \cite{Fix:2003pc} (see also 
     \cite{Fix:2001cz}).}
\end{figure}

As a general trait in meson production in nucleon--nucleon scattering, the 
primary production amplitude, i.\ e.\ the underlying dynamics can be regarded 
as energy independent in the vicinity of threshold \cite{Moalem:1996ua,
Bernard:1998sz,Gedalin:1998zk}.
Consequently, for s--wave production processes, the energy dependence of the 
total cross section is essentially given by a phase space behaviour modified 
by the influence of final state interactions.
In Fig.\ref{fig_eta_etap} total cross section data obtained in proton--proton 
scattering are shown for the pseudoscalar isosinglet mesons $\eta$ and 
$\eta^\prime$ \cite{Moskal:2003gt}.
In both cases, the energy dependence of the total cross section deviates 
significantly from phase space expectations.
Including the on--shell $\mbox{}^1\mbox{S}_0$ proton--proton FSI enhances the 
cross section close to threshold by more than an order of magnitude, in good 
agreement with data in case of $\eta^\prime$.
As expected from kinematical considerations \cite{Moskal:2002jm} the cross 
section for $\eta$ production deviates from phase space including the $pp$ FSI 
at excess energies $\mbox{Q} \ge 40\,\mbox{MeV}$, where the 
$\mbox{}^1\mbox{S}_0$ final state is no longer dominant compared to higher 
partial waves.
Deviations at low excess energies seem to be well accounted for by an 
attractive proton--$\eta$ FSI (dashed line), treated phenomenologically as an 
incoherent pairwise interaction \cite{Bernard:1998sz,Moskal:2002jm,
Schuberth:1995pd}.
%\footnote{Values of proton--$\eta$ scattering length $\mbox{a}_{p \eta} = 
%(0.7 + i\,0.4)\,\mbox{fm}$ and effective range $\mbox{r}_{p \eta} = 
%- (1.50 + i\,0.24)\,\mbox{fm}$ are taken from \cite{Green:1997yi}.}
In comparison to the proton--$\eta^\prime$ (Fig.\ref{fig_eta_etap}) and 
proton--$\pi^0$ systems only the $p \eta$ interaction is strong enough to 
become apparent in the energy dependence of the total cross section 
\cite{Moskal:2000pu}.
In differential observables, effects should be more pronounced in the phase 
space region of low proton--$\eta$ invariant masses.
However, to discern effects of proton--$\eta$ scattering from the influence of 
proton--proton FSI, which is stronger by two orders of magnitude, requires high 
statistics measurements, which have only become available recently 
\cite{Calen:1998kx,Abdel-Bary:2002sx,Moskal:2003gt}:
\begin{figure}[htb]
  \label{fig_eta_inv_pp}
  \includegraphics[height=.32\textheight]{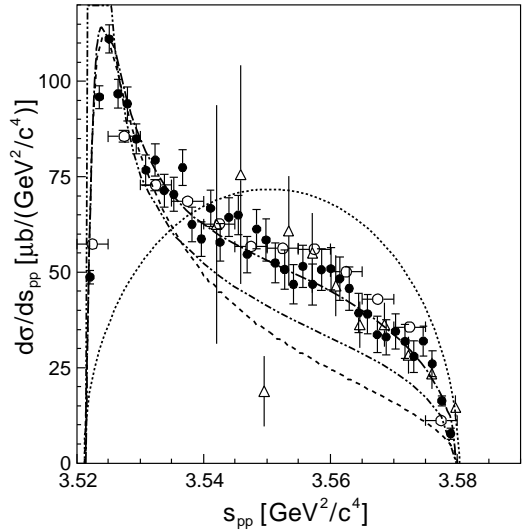}
  \caption{Invariant mass squared of the ($pp$)--subsystem in the reaction 
     $pp \rightarrow p p \eta$ at excess energies of 
     $\mbox{Q} = 15.5\,\mbox{MeV}$ (COSY--11, solid circles 
     \cite{Moskal:2003gt}, $\mbox{Q} = 15\,\mbox{MeV}$ (COSY--TOF, open 
     circles \cite{Abdel-Bary:2002sx} and $\mbox{Q} = 16\,\mbox{MeV}$ 
     (PROMICE/WASA, open triangles \cite{Calen:1998kx}). The dotted and dashed 
     lines follow a pure phase space behaviour and its modification by the 
     phenomenological treatment of the three--body FSI as an incoherent 
     pairwise interaction, respectively. The latter was normalized at small invariant 
     mass values. Effects from including a P--wave admixture in the $pp$ 
     system are depicted by the dashed--dotted line \cite{Nakayama:2003jn}, 
     while the dashed--double--dotted line corresponds to a pure s--wave final 
     state with a full three--body treatment \cite{Fix:2001cz}.}
\end{figure}
Close to threshold, the distribution of the invariant mass of the 
proton--proton subsystem is characteristically shifted towards low invariant 
masses compared to phase space (dotted line in Fig.\ref{fig_eta_inv_pp}).
This low--energy enhancement is well reproduced by modifying phase space with 
the $\mbox{}^1\mbox{S}_0$ $pp$ on--shell interaction. 
A second enhancement at higher $pp$ invariant masses, i.e.\ low energy in the 
$p \eta$ system, is not accounted for even when including additionally the 
proton--$\eta$ interaction incoherently (dashed line).
However, including a P--wave admixture in the $pp$ system by considering a 
$\mbox{}^1\mbox{S}_0 \rightarrow \mbox{}^3\mbox{P}_0 \mbox{s}$ transition in 
addition to the $\mbox{}^3\mbox{P}_0 \rightarrow \mbox{}^1\mbox{S}_0 \mbox{s}$ 
threshold amplitude, excellent agreement with the experimental invariant mass 
distribution is obtained (dashed--dotted line \cite{Nakayama:2003jn}).
In return, with the P--wave strength adjusted to fit the invariant mass data, 
the approach fails to reproduce the energy dependence of the total cross 
section (Fig.\ref{fig_eta_etap}) below excess energies of 
$\mbox{Q} = 40\,\mbox{MeV}$.
Preliminary calculations considering only s--waves in the final state but using 
a rigorous three--body treatment of the $pp \eta$ final state actually decrease 
the cross section at large values of the $pp$ invariant mass 
(dashed--double--dotted lines \cite{Fix:2001cz}) compared to an incoherent 
two--body calculation within the same framework.
However, close to threshold the energy dependence of the total cross section is 
enhanced compared to the phenomenological incoherent treatment and the data 
(Fig.\ref{fig_eta_etap}).
Although part of this enhancement has to be attributed to the neglect of 
Coulomb repulsion in the $pp$ system, consequently overestimating the $pp$ 
invariant mass at low values, qualitatively the full three--body treatment has 
opposite effects compared to a P--wave admixture in the proton--proton system 
in view of both the total cross section as well as the $pp$ invariant mass 
distribution. 
In the approximate description of the total cross section by the phenomenological 
s--wave approach with an incoherent FSI treatment these two effects seem to 
cancel casually.

Close to threshold, resonance excitation of the $\mbox{S}_{11}\,(1535)$ and 
subsequent decay to the $p \eta$ final state is generally\footnote{The $pn 
\rightarrow d \eta$ excitation function has been interpreted to provide direct 
experimental evidence for $\mbox{S}_{11}\,(1535)$ excitation 
\cite{Calen:1997sf}. It should be noted, however, that in \cite{Pena:2000gb} 
for $\eta$ production in proton--proton scattering short range nucleonic 
currents, i.e.\ $\sigma$ and $\omega$ exchange are found to be much stronger 
compared to the contribution from resonance currents.} believed to be the 
dominant $\eta$ production mechanism \cite{Batinic:1997me,Santra:1998jf,
Gedalin:1998fa,Bernard:1998sz,Faldt:2001uz,Nakayama:2002mu,Baru:2002rs,
Nakayama:2003jn}.
In this context, the issue of the actual excitation mechanism of the 
$\mbox{S}_{11}\,(1535)$ remains to be addressed.
The $\eta$ angular distribution is sensitive to the underlying dynamics: 
A dominant $\rho$ exchange favoured in \cite{Faldt:2001uz} results in an 
inverted curvature of the $\eta$ angular distribution compared to $\pi$ and 
$\eta$ exchanges which are inferred to give the largest contribution to 
resonance excitation in \cite{Nakayama:2002mu}.
In the latter approach the interference of the pseudoscalar exchanges in the 
resonance current with non--resonant nucleonic and mesonic exchange currents 
turns the curvature to the same angular dependence as expected for $\rho$ 
exchange.
Presently, due to the statistical errors of the available unpolarised data 
at an excess energy of $\mbox{Q} \approx 40\,\mbox{MeV}$ 
\cite{Calen:1998kx,Abdel-Bary:2002sx} it is not possible to differentiate 
between a dominant $\rho$ or $\pi$, $\eta$ exchange, as discussed in 
\cite{Abdel-Bary:2002sx}.
Data recently taken at the CELSIUS/WASA facility with statistics increased by 
an order of magnitude compared to the available data might provide an answer in 
the near future \cite{Zlomanczuk:2003pc}.

Spin observables, like the $\eta$ analyzing power, should even disentangle a 
dominant $\rho$ meson exchange and the interference of $\pi$ and $\eta$ 
exchanges in resonance excitation with small nucleonic and mesonic currents 
\cite{Nakayama:2002mu}, which result in identical predictions for the 
unpolarised $\eta$ angular distribution.
First data \cite{Winter:2002ft} seem to favour the vector dominance model, but 
final conclusions both on the underlying reaction dynamics and the admixture of 
higher partial waves \cite{Nakayama:2003jn} have to await the analysis of data 
taken with higher statistics for the energy dependence of the $\eta$ analysing 
power \cite{Czyzykiewicz:2003ha}.

\section{Associated Strangeness Production}
In elementary hadronic interactions with no strange valence quark in the 
initial state the associated strangeness production provides a powerful tool to 
study reaction dynamics by introducing a ``tracer'' to hadronic matter.
Thus, quark model concepts might eventually be related to mesonic or baryonic 
degrees of freedom, with the onset of quark degrees of freedom expected for 
kinematical situations with large enough transverse momentum transfer.

First exclusive close--to--threshold data on $\Lambda$ and $\Sigma^0$ production 
\cite{Balewski:1998pd,Sewerin:1998ky} obtained at the COSY--11 facility showed 
at equal excess energies below $\mbox{Q} = 13\,\mbox{MeV}$ a cross section 
ratio of 
\begin{equation}
{\cal{R}}_{\Lambda/\Sigma^0} \, \left(\mbox{Q} \le 13\,\mbox{MeV}\right) = 
\frac{\sigma\left(pp \rightarrow pK^+\Lambda \right)}
     {\sigma\left(pp \rightarrow pK^+\Sigma^0 \right)} =
28^{+6}_{-9}
\label{eq_ls_ratio}
\end{equation}
exceeding the high energy value ($\mbox{Q} \ge 300\,\mbox{MeV}$) of 2.5 
\cite{Baldini:1988th} by an order of magnitude.

%In the meson exchange framework, considering only $\pi$ exchange, data on $\pi$ 
%induced hyperon production via $\pi N \rightarrow 
%K \Lambda\left(\Sigma^0\right)$ result in a ratio of 
%${\cal{R}}_{\Lambda/\Sigma^0} \approx 0.9$ \cite{Gasparian:1999jj}, clearly 
%underestimating the experimental value \eqref{eq_ls_ratio}.
%Kaon exchange essentially relates the ratio ${\cal{R}}_{\Lambda/\Sigma^0}$ to 
%the ratio of coupling constants at the nucleon--hyperon--kaon vertices 
%$\mbox{g}_{NYK}$ squared, which are not well known experimentally.
%The suitable choice of the SU(6) prediction \cite{Dover:1985ba,deSwart:1963gc} 
%leads to a $\Lambda/\Sigma^0$ production ratio of 27 for pure $K$ exchange, in 
%good agreement with the experimental value.
%However, effects of final state interaction and the importance of $\pi$ 
%exchange for $\Sigma$ production are neglected by this simple estimate.
In the meson exchange framework, estimates for $\pi$ and $K$ exchange 
contributions based on the elementary scattering processes do not reproduce the 
experimental value \eqref{eq_ls_ratio} \cite{Sewerin:1998ky,Gasparian:1999jj}.
However, inclusive $K^+$ production data in $pp$ scattering at an excess energy 
of $\mbox{Q} = 252\,\mbox{MeV}$ with respect to the $p K^+ \Lambda$ threshold 
show enhancements at the $\Lambda p$ and $\Sigma N$ thresholds of similar 
magnitude \cite{Siebert:1994jy}.
Qualitatively, a strong $\Sigma^0 N \rightarrow \Lambda p$ final state 
conversion might account for both the inclusive SATURNE results as well as the 
$\Sigma^0$ depletion in the COSY--11 data.
Evidence for such conversion effects is known e.\ g.\ from fully constrained 
kaon absorption on deuterium via $K^- d \rightarrow \pi^- \Lambda p$ 
\cite{Tan:1969jq}.

In exploratory calculations performed within the framework of the J\"ulich meson 
exchange model \cite{Gasparian:1999jj}, taking into account both $\pi$ and $K$ 
exchange diagrams in a coupled channel approach, a final state conversion is 
rather excluded as origin of the experimentally observed ratio:
While $\Lambda$ production is found to be dominated by kaon exchange, both 
$\pi$ and $K$ exchange turn out to contribute to the $\Sigma^0$ channel with 
similar strength.
Qualitatively, this result is experimentally confirmed at higher excess 
energies between $\mbox{Q} = 200\,\mbox{MeV}$ and $\mbox{Q} = 430\,\mbox{MeV}$ 
from polarization transfer measurements from the DISTO experiment 
\cite{Balestra:1999br,Maggiora:2002nm,Laget:1991jk}.
It is concluded in \cite{Gasparian:1999jj}, that only a destructive interference 
of $\pi$ and $K$ exchange might explain the experimental value 
\eqref{eq_ls_ratio}.
$\Sigma$ production in different isospin configurations should provide a 
crucial test for this interpretation, since for the reaction $pp \rightarrow 
n K^+ \Sigma^+$ the interference pattern is found to be opposite compared to 
the $p K^+ \Sigma^0$ channel.
Data close to threshold have recently been taken at the COSY--11 
facility \cite{Rozek:2002}.

However, within an effective Lagrangian approach \cite{Shyam:2000sn} both 
$\Lambda$ and $\Sigma^0$ production channels are concluded to be dominated by 
$\pi$ exchange and excitation of the $\mbox{S}_{11}\,(1650)$ close to threshold, 
while at excess energies above $\mbox{Q} = 300\,\mbox{MeV}$ the $N^*(1710)$ 
governs strangeness production\footnote{For further complementary theoretical 
approaches see references in \cite{Moskal:2002jm,Kowina:2003ha,Wagner:2003ha}.}.
In this energy range the influence of resonances becomes evident from recent 
data on invariant mass distribution determined at COSY--TOF \cite{Wagner:2003ha}.

To study the transition region between the low--energy enhancement 
\eqref{eq_ls_ratio} and the high energy value measurements have been extended up 
to excess energies of $\mbox{Q} = 60\,\mbox{MeV}$ 
\cite{Kowina:2003ha,Kowina:2003kx}:
%The energy dependence of the total cross section for $\Lambda$ production is 
%much better described by a phase space behaviour modified by the $p \Lambda$ 
%final state interaction than by pure phase space \cite{Kowina:2003ha}.
%In contrast, $\Sigma^0$ production is equally well described neglecting any FSI 
%effect.
In order to describe the energy dependence of the total cross section for 
$\Lambda$ production, in addition to phase space the $p \Lambda$ final state 
interaction has to be taken into account.
In contrast, $\Sigma^0$ production is satisfactorily well described by phase 
space behaviour only \cite{Kowina:2003ha}.
This qualitatively different behaviour might be explained by the $\Sigma^0 p$ 
FSI being much weaker compared to the $\Lambda p$ system.
However, the interpretation implies dominant S--wave production and reaction 
dynamics that can be regarded as energy independent.
Within the present level of statistics, contributions from higher partial waves 
can be neither ruled out nor confirmed at higher excess energies for $\Sigma^0$ 
production.

\begin{figure}[htb]
  \label{fig_ls_ratio}
  \includegraphics[height=.28\textheight]{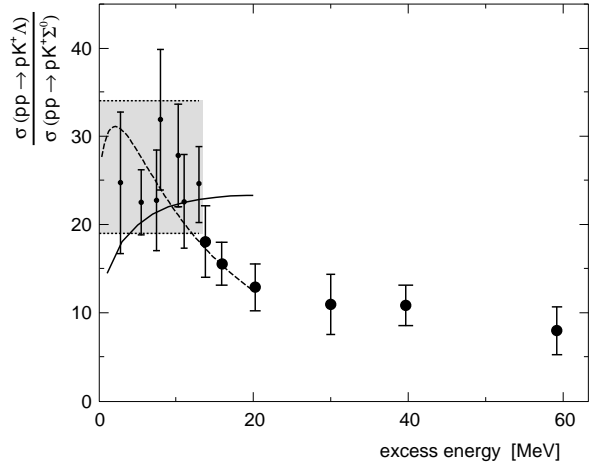}
  \caption{$\Lambda/\Sigma^0$ production ratio in proton--proton scattering as 
     a function of the excess energy. Data are from \cite{Sewerin:1998ky} 
     (shaded area) and \cite{Kowina:2003kx}. Calculations 
     \cite{Gasparyan:2002sy} within the J\"ulich meson exchange model imply a 
     destructive interference of $K$ and $\pi$ exchange using the microscopic 
     Nijmegen NSC89 (dashed line \cite{Maessen:1989sx}) and the new J\"ulich 
     model (solid line \cite{Haidenbauer:2001gx}) for the $Y N$ final state 
     interaction.}
\end{figure}

The energy dependence of the production ratio ${\cal{R}}_{\Lambda/\Sigma^0}$ is 
shown in Fig.\ref{fig_ls_ratio} in comparison with theoretical calculations 
obtained within the approach of \cite{Gasparian:1999jj} assuming a destructive  
interference of $\pi$ and $K$ exchange and employing different choices of the 
microscopic hyperon nucleon model to describe the interaction in the final 
state \cite{Gasparyan:2002sy}.  
The result crucially depends on the details --- especially the off--shell 
properties --- of the hyperon--nucleon interaction employed.
At the present stage both the good agreement found in \cite{Gasparian:1999jj} 
with the threshold enhancement \eqref{eq_ls_ratio} and for the Nijmegen model 
(dashed line in Fig.\ref{fig_ls_ratio}) with the energy dependence of the cross 
section ratio should rather be regarded as 
accidental\footnote{In the latter case an SU(2) breaking in the 
$\mbox{}^3\mbox{S}_1$ $\Sigma N$ channel had to be introduced 
\cite{Maessen:1989sx} resulting in an ambiguity for the $\Sigma^0 p$ amplitude.}.
Calculations using the new J\"ulich model (solid line in Fig.\ref{fig_ls_ratio}) 
do not reproduce the tendency of the experimental data.
It is suggested in \cite{Gasparyan:2002sy} that neglecting the energy dependence 
of the elementary amplitudes and higher partial waves might no longer be 
justified beyond excess energies of $\mbox{Q} = 20\,\mbox{MeV}$.
However, once the reaction mechanism for close--to--threshold hyperon production 
is understood, exclusive data should provide a strong constraint on the details 
of hyperon--nucleon interaction models.

\section{Present and Future}
Intermediate baryon resonances emerge as a common feature in the dynamics of the 
exemplary cases for threshold meson production in nucleon--nucleon scattering 
discussed in this article. 
However, this does not hold in general for meson production in the 
$1\,\mbox{GeV}/\mbox{c}^2$ mass range (for a discussion on $\eta^\prime$ 
production see \cite{Moskal:2003ha}).
Moreover, the extent to which resonances are evident in the observables or 
actually govern the reaction mechanism depends on the specific channels, which 
differ in view of the level of present experimental and theoretical 
understanding.

The $N^*(1440)$ resonance dominates $\pi^+ \pi^-$ production at threshold, and 
exclusive data allow to extract resonance decay properties in the low--energy 
tail of the Roper.
Dynamical differences between the different isospin configurations of the 
$\pi \pi$ system and the behaviour at higher energies remains to be understood 
with first experimental clues appearing.

With three strongly interacting particles in the final state, a consistent 
description of $\eta$ production close to threshold requires an accurate 
three--body approach taking into account the possible influence of higher 
partial waves. 
High statistics differential cross sections and polarization observables coming 
up should straighten out both the excitation mechanism of the $N^*(1535)$ and 
the admixture of higher partial waves.

At present, the available experimental data on the elementary strangeness 
production channels give evidence for both an important role of resonances 
coupling to the hyperon--kaon channels and on a dominant non--resonant kaon 
exchange mechanism.
Experiments on different isospin configurations, high statistics and spin 
transfer measurements close to threshold should disentangle the situation in 
future. 

From the cornerstone of total cross section measurements, it is apparent from 
the above examples to what extent our knowledge is presently enlarged by 
differential observables and what will be the impact of polarization experiments 
in future to get new perspectives in threshold meson production.

%%%%%%%%%%%%%%%%%%%%%%%%%%%%%%%%%%%%%%%%%%%%%%%%
%% BACKMATTER
%%%%%%%%%%%%%%%%%%%%%%%%%%%%%%%%%%%%%%%%%%%%%%%%

\begin{theacknowledgments}
The author gratefully acknowledges the pleasure to work with the CELSIUS/WASA 
and COSY--11 collaborations, and, in particular, thanks M.\ Bashkanov, 
H.\ Clement, R.\ Meier, P.\ Moskal and W.\ Oelert for helpful discussions .
This work has been supported by The Swedish Foundation for International 
Cooperation in Research and Higher Education (STINT Kontrakt Dnr 02/192). 
\end{theacknowledgments}

%% If the bibliography is produced without BibTeX comment out the

\bibliographystyle{aipproc}   % if natbib is available

\bibliography{had03_abbrev,had03_mw}

\IfFileExists{\jobname.bbl}{}
 {\typeout{}
  \typeout{******************************************}
  \typeout{** Please run "bibtex \jobname" to optain}
  \typeout{** the bibliography and then re-run LaTeX}
  \typeout{** twice to fix the references!}
  \typeout{******************************************}
  \typeout{}
 }

\end{document}